**Relationship between Periodontal disease and Asthma among overweight/obese adults**

Author Information:

**Roberto Rivera, PhD, College of Business, University of Puerto Rico, Mayaguez, [roberto.rivera30@upr.edu](mailto:roberto.rivera30@upr.edu) (corresponding author)**
Oelisoa M. Andriankaja, PhD, School of Dental Medicine, University of Puerto Rico, Medical Science Campus
Cynthia M. Perez, PhD, School of Public Health, University of Puerto Rico, Medical Science Campus
Kaumudi Joshipura, ScD, School of Dental Medicine, University of Puerto Rico, Medical Science Campus

This research has been supported by NIH/NIDCR (R01DE020111)

**Declaration of interest**

The authors report no conflicts of interest

**Abstract**

**Aim:** To assess the relationship between oral health and asthma.

**Methods**: Data from 1,315 overweight or obese individuals, aged 40-65 years was used. Asthma was self-reported, while periodontitis, bleeding on probing (BOP), and plaque index were determined by clinical examinations.

**Results**: Using logistic regression adjusting for gender, smoking status, age, body mass index, family history of asthma, and income level, revealed that the odds ratio (OR) of asthma for a participant with severe periodontitis was 0.44 (95% confidence interval: 0.27, 0.70) that of a participant with none/mild periodontitis . On the other hand, proportion of BOP sites, and plaque index were not statistically significant. For a participant with severe periodontitis, the OR of taking asthma medication was 0.20 (95% confidence interval: 0.09, 0.43) that of a participant with none/mild periodontitis.  Moreover, proportion of BOP sites was statistically associated to use of asthma medication while plaque index still remained non-significant.

**Conclusion**: Participants with severe periodontitis were less likely to have asthma. Stronger evidence of an inverse association was found when using asthma medication as outcome.

Keywords: asthma; periodontal disease; asthma medication; periodontitis; hygiene hypothesis

**Clinical Relevance**

*Scientific rationale for study:* Asthma is an increasingly serious problem around the world. Asthma may be an adverse effect of less exposure to bacteria and microbes, perhaps through modern oral hygiene routines.

*Principal findings:* Severe periodontitis and proportion of sites bleeding on probing were inversely associated with asthma, supporting the hygiene hypothesis.

Practical implications: The association between asthma and oral health can potentially lead to novel asthma prevention strategies.

## Introduction

*Rates of periodontal disease and tooth decay are generally down partly due to advancements in dentistry (Arbes & Matsui 2011, Dye et al. 2007) and partly due to a reduction of risk factors. Researchers have wondered what type of association advancements on dentistry have with other diseases, including asthma and allergy-related illnesses.*

*Arbes and Matsui (2011) examined 12 studies evaluating the association between allergic disease or asthma and oral bacteria or periodontal disease. Five found an inverse association between allergic disease or asthma and periodontal health, four showed a positive association, and three found no link. A recent small study also found a positive association between periodontal disease and severity of asthma (Gomes-Filho et al. 2013). Inverse associations have been attributed to the hygiene hypothesis. Through this hypothesis it is argued that with better oral health, fewer opportunities for infections and microbial exposures exist. As a consequence, asthma and atopic diseases become more likely (Arbes et al. 2006). Positive associations, on the other hand, have been attributed to mouth breathing habits and frequent use of inhalational drugs in patients suffering from severe asthma (Gomes-Filho et al. 2013, Ramesh-Reddy et al. 2014). The inconsistent results of the studies may be partially attributed to the use of different definitions of periodontal disease. In 2007, the Centers for Disease Control and Prevention and the American Academy of Periodontology (CDC-AAP) developed standard case definitions for periodontitis for use in surveillance and population-based research(Page & Eke 2007).*

*Although some studies evaluating the association between oral pathogens and allergy-related outcomes have considered the possible confounding effects of ethnicity or race, there is*

*evidence of differences in morbidity of these diseases among Hispanics. Specifically, Puerto Ricans have been found to have higher asthma morbidity and mortality rates (Bartolomei-Diaz et al. 2009, Choudhry et al. 2005) and higher severe periodontitis than other Latino groups (Jimenez et al. 2014).*

*The objective of this study was to determine if there is an association between periodontal disease and asthma. This study used baseline data from the ongoing San Juan Overweight Adults Longitudinal Study (SOALS).*

## Methods

### Study population:

*SOALS is a three-year follow up study of overweight or obese individuals aged 40 to 65 years, residing in the San Juan municipality and its vicinity in Puerto Rico. Participants were excluded if they: 1) had less than four natural teeth or had braces or orthodontic appliances to ensure the validity of periodontal measurements; 2) were pregnant; 3) had been diagnosed with health conditions which could potentially increase the risk of systemic complications from the periodontal exam, such as: congenital heart murmurs, heart valve disease, congenital heart disease, endocarditis, stroke, rheumatic fever, and hemophilia or bleeding disorders; 4) had active dialysis treatment; 5) had active anticoagulant therapy; 6) were prescribed antibiotic prophylaxis prior to surgical or dental procedures; 7) had undergone procedures related to cardiovascular disease including pacemaker, defibrillator implantation, or surgery on the heart or vessels that involves use of prosthetic material; 8) had undergone hipbone or other joint replacement surgery; or 9) planned on moving away in the next three-year period. Although participants previously diagnosed with diabetes were excluded, provisional diagnosis of diabetes was determined based on the American Diabetes Association cutoffs.*

Of the adults that responded to the study advertisement, 1,931 adults were eligible from the screening interview. A total of 1,610 adults were scheduled for the baseline visit, of which 1,455 underwent the study procedures. From these, a total of 61 participants were ineligible based on criteria defined above. Accounting for exclusions, 1,394 participants were eligible to participate in the study and complete data was available for 1,315. The Institutional Review Board (IRB) at the University of Puerto Rico approved the study and all participants gave written consent prior to inclusion in the study.

***Definition of asthma:***

Questionnaire data was used to define asthma. The first definition was self-reported physician-diagnosed asthma, which was based on the question "Have you ever been diagnosed with asthma?" The second definition was based on both, physician-diagnosed asthma and a positive response to the question "Have you ever taken any medications or treatment for asthma?" This second definition should help to better distinguish between those who have moderate to serious asthma and those who have mild asthma. Participants were also queried about immediate family's history of asthma.

***Periodontal examination and periodontitis definition:***

The periodontal examination consisted of measures of probing depth, gingival recession, plaque index, bleeding upon probing (BOP) and computed clinical attachment loss, all performed by trained dental examiners. The measurements are described in a previous publication (Andriankaja & Joshipura 2014). The CDC-AAP periodontitis definition was used to categorize periodontitis (Page & Eke 2007). This definition uses both attachment loss and probing depth measurements to determine periodontitis severity. Specifically, periodontitis is classified as severe (at least two interproximal sites with attachment loss of 6 mm or greater [not on same tooth] or at least one interproximal site with pocket depth of 5mm or greater), moderate (at least two interproximal sites with attachment loss of 4 mm or greater [not on same tooth] or at least two interproximal site with pocket depth of 5mm or greater [not on same tooth]), or none/mild (neither moderate nor severe periodontitis).

BOP, a surrogate measure of current oral clinical inflammation which could be related to a mix of reversible gingivitis due to poor oral hygiene status and the actual deep pocket inflammation (periodontitis), was determined by the presence or absence of bleeding at any of the six tooth sites throughout the periodontal probing measurement. Hence, BOP was noted by gently inserting the probe to the base of the sulcus or pocket probing force of no more than 20 g: BOP was present if the site bled within 20s following the probing (Chaves et al. 1993). The extent and severity of the BOP was defined by the proportion of sites with bleeding upon probing. The Silness and Loe Plaque Index, a proxy measure of oral hygiene status, was determined by visual assessment of any presence of plaque after passing a probe around the tooth surface of six pre-selected Ramfjord teeth (Fleiss et al. 1987). Plaque index was coded as: 0: Absence of plaque; 1: Presence of plaque following the passage of the probe around the tooth; 2: Visible plaque; 3: Abundant plaque.  For each participant the mean plaque index over all sites and the proportion of BOP (PBOP) sites was calculated. The reliability-validity analysis of the dental probing showed a 96% concordance within 1mm of clinical attachment loss between the dental examiners and the National Health and Nutrition Examination Survey (NHANES) chief examiner.

*Other measures:*

Information on covariates including demographic characteristics were obtained from the questionnaire completed during the baseline visit. Participants were categorized as never smokers if they reported smoking at most 100 cigarettes during their entire life, otherwise they were classified as current smokers. Anthropometric measures, were taken at least twice based on the NHANES III anthropometric procedures, and the mean of the measures was used (CDC 2008). Weight was measured to the closest 0.5 kg and height to the nearest 0.1 cm. The body mass index (BMI) was calculated as weight in kilograms divided by the height in square meters ($kg/m^2$).

*Statistical analysis*

We tested the association between each asthma definition and periodontal disease exposures through logistic regression. Significance level was set to 5%. Interpretations rely on calculated Odds Ratios (OR). Logistic regression models were adjusted for age, gender, smoking status, BMI, family history of asthma, and income level. Since logistic regression assumes relationships are strictly linear, local polynomial regression (Cleveland et al. 1992) were plotted to explore the possibility of nonlinear associations. When these indicated a possible nonlinear association, regression splines were fit to draw inference (Wood 2011). However, the regression spline model results did not suggest nonlinearity and hence, are not presented. Our outcome of interest was asthma using the definitions above.

## Results

The characteristics of participants with and without asthma were similar with regards to age, annual household income, and BMI (Table 1). However, a larger proportion of males were observed among participants without asthma, whereas there were more current smokers among participants with asthma. The percentage of participants with none/mild or moderate periodontitis was similar among participants regardless of asthma prevalence reported. However, the proportion of persons with severe periodontitis was much lower among participants who reported asthma in comparison to those who did not. Similarly, median proportion BOP and mean plaque index were also lower for participants with asthma than those

*without, especially for those taking asthma medication. There was no difference in median number of missing teeth in each asthma group.*

*Overall, participants with severe periodontitis were less likely to be diagnosed with asthma, but there is little difference in odds of asthma for participants diagnosed with moderate periodontitis or no periodontitis (Table 2). Furthermore, the odds of asthma increased with BMI, for smokers, and for those with a family history of asthma. For a participant with severe periodontitis, the odds ratio (OR) of asthma was 0.44 (95% confidence interval: 0.27, 0.70) that of a participant with none/mild periodontitis (Table 2). On the other hand, the association between PBOP and plaque index with asthma were not statistically significant. No evidence was found of any interaction effects between the exposures and the confounding variables.*

*Results using the second definition of asthma also indicate an inverse association between use of asthma medication and severity of periodontitis (Table 2). Specifically, for a participant with severe periodontitis, the OR of taking asthma medication was 0.20 that of a participant with none/mild periodontitis (95% CI: 0.09, 0.43). Moreover, PBOP was strongly associated with asthma medication while plaque index still remained non-significant. Our results indicate that a 0.1 increase in proportion of BOP decreases odds of taking asthma medication by about 13% (=$0.26^{0.1}$). Compared to the asthma prevalence model, participants with severe periodontitis had an even lower estimated probability of taking asthma medicine, and a slight difference in probability of taking asthma medicine for participants diagnosed with moderate periodontitis versus none/mild periodontitis. As we can see from Figure 1, the probability of taking asthma medicine was inversely associated with PBOP. Furthermore, similar to the asthma diagnosis*

*model, the odds of taking asthma medicine increased with BMI and for smokers, while decreasing for participants with no history of family asthma prevalence.*

**Discussion**

Our study demonstrated that, on average, participants with severe periodontitis or higher PBOP are less likely to have asthma. Severe periodontitis resulted in the largest reduction on the likelihood of asthma. Stronger evidence of an inverse association between asthma and oral health was found when asthma medicine intake was used as an indicator of asthma severity instead of asthma diagnosis. The results of this analysis support the hygiene hypothesis (Strachan 1989). Five other studies found an inverse association between asthma or allergy and periodontal disease (Arbes et al. 2006, Friedrich et al. 2008, Friedrich et al. 2006, Grossi et al. 1994, Hujoel et al. 2008). Several other studies either did not find an association between allergy or asthma and periodontal disease variables in humans (
Abrahamsson et al. , Eloot et al. 2004, Shulman et al. 2003, von Hertzen L. C. et al. 2006) or found a positive association between these variables (Gomes-Filho et al. 2013, Laurikainen & Kuusisto 1998, McDerra et al. 1998, Mehta et al. 2009, Stensson et al. 2010). The diverging conclusions in the literature can be attributed to several factors. Most noticeably, both asthma and periodontitis are affected by multiple factors. Some authors argue that mouth breathing habits and frequent use of inhalational drugs in patients suffering from severe asthma may lead to a positive association between asthma severity and periodontal disease, contrary to our findings (Gomes-Filho et al. 2013, Ramesh-Reddy et al. 2014). Furthermore, periodontal disease was measured in different ways in these studies. For example, Gomes-Filho *et al.*, (2013) defined periodontitis based on clinical measures, but did not distinguish between none/mild, moderate and severe periodontitis, only using a yes/no periodontitis definition. Moreover, their periodontitis case definition included bleeding on probing, and compared participants with no asthma against patients with severe asthma. On the other hand, Friedrich et al. (2008) used attachment loss but not pocket depth to define periodontitis severity. Since attachment loss can accompany non-inflammatory gingival recession, determining periodontitis severity using attachment loss alone may overstate severity (Page & Eke 2007). Several studies suffer from small sample sizes, and limited control of confounding variables. Another factor was the way asthma diagnosis was determined. In most studies, asthma diagnosis was self-reported, based on whether they have been diagnosed with asthma or allergies by a health professional . None of the studies used challenge tests to determine asthma diagnosis. Aaron et al. (2008) found that about a third of patients diagnosed with asthma by a physician did not have asthma when assessed through lung function and challenge tests.

We did not find any association between mean plaque index and asthma prevalence or taking asthma medication. A possible explanation is that although the Silness Loe plaque index reflects the oral hygiene status and expresses the presence of plaque at or above the gum line, it does not assess the amount of plaque below the gum line, or the bacterial composition of plaque which may be more pertinent for periodontitis, and perhaps asthma.

Our study, used the CDC-AAP defined periodontitis based on clinical periodontal examination of participants (Page & Eke 2007). By using both attachment loss and probing depth measurements, this definition better reflects the cumulative and recent impact of periodontitis and its treatment, in comparison to using attachment loss or probing depth alone.  A more recent CDC-AAP definition distinguishes between no periodontitis and mild periodontitis (Eke et al. 2012). However, the most recent definition did not change the overall interpretation of the statistical analysis conducted in this work. Furthermore, although we use a two category smoking status, in our preliminary analysis using an additional 'former smoker' category did not alter our results. The self-reported measure of smoking status may not be precise enough to account for the true influence of tobacco exposure. Other measures of tobacco exposure such as toenail nicotine levels (Al-Delaimy & Willett 2008) may be more useful. To our knowledge, this is the first study assessing the association between periodontal disease and asthma on Puerto Ricans, a group disproportionately affected by both asthma and severe periodontitis. Our results have limitations. Asthma prevalence was self-reported and does not provide indication of severity of the condition as provided by lung function and challenge tests. Not enough information is available from the study to determine the adherence to asthma control medication. Findings were for an overweight or obese adult population. Obese asthma patients may represent a different asthma phenotype (Lugogo et al. 2010). The cross-sectional nature of our association analysis does not prove causation and the causal relationship may be in the opposite direction. It is also plausible that the causal relationship is bi-directional (Garlet 2010). Another limitation is that a convenience sample was used, limiting generalizability.

Several animal studies support the hygiene hypothesis' principle of a causal relationship between a decline of infectious diseases and increase in immunological disorders (Card et al. 2010, Navarro et al. 2011, Okada et al. 2010, Olszak et al. 2012, Versini et al. 2015). Unfortunately, longitudinal studies involving young children to assess the association between asthma or allergic diseases and directly measured oral bacteria currently do not exist.

Recently, results from a birth cohort study indicate that early life exposure to certain allergens and bacteria might protect from recurrent wheezing and allergic diseases (Lynch et al. 2014). One biological mechanism that might explain the hygiene hypothesis is that with less exposure to microbes during early childhood, a shift from a $T_H2$ to a $T_H1$ phenotype is hindered

(Romagnani 2007). Recently, studies postulated that an imbalance in $T_H1$ and $T_H2$ associated chemokines may precede the onset of sensitization, eczema and recurrent wheeze from birth, implicating that these chemokines may sometimes be primarily involved in the pathogenesis of allergic diseases and not only secondary to a general immune deviation after disease onset ( Abrahamsson et al. ). With the discovery of regulatory T cells, a second mechanism was proposed: reduced microbial exposure leads to reduced $T_H$ cell suppression by regulatory T cells. Ramesh-Reddy et al. (2014) argue that microbial exposure early in life induces $T_H1$ cell mediated response and promotes regulatory T cell network that later suppresses T cell immune responses, including $T_H2$ cell mediated respiratory allergic diseases. Still, the complexity of the microbial and immunological interactions implies a high number of asthmatic immune phenotypes (Hansel et al. 2013).

Novel allergic disease prevention strategies currently being studied include colonizing patients with bacteria or other microbes, modifying helminth exposure, and molding gut microbial diversity (Arbes & Matsui 2011, Smits et al. 2016). Identification of commensal oral bacteria which provide protection from allergic diseases is still needed. For example an experimental study found that Porphyromonas gingivalis (P. gingivalis) infection before allergy sensitization reduced airway inflammation but had no effect on airway hyperresponsiveness to a methacholine aerosol challenge (Card et al. 2010). On the other hand, P. gingivalis infection established after allergic sensitization reduced airway hyperresponsiveness but did not alter inflammation. Given that the hygiene hypothesis addresses exposure to bacteria and other microbes, further research is needed to better elucidate the association between periodontal disease and asthma. In addition, case control or prospective cohort studies to assess the relationship between oral bacteria and allergic disease in young children are also needed.

periodontitis', *Community dentistry and oral epidemiology* **15**(4), 221–224. Epub 1987/08/01. PubMed PMID: 3476248.

Friedrich, N., Kocher, T., Wallaschofski, H., Schwahn, C., Ludemann, J., Kerner, W. & et al. (2008), 'Inverse association between periodontitis and respiratory allergies in patients with type 1 diabetes mellitus', *J Clin Periodontol* **35**, 305–310.

Friedrich, N., Volzke, H., Schwahn, C., Kramer, A., Junger, M., Schafer, T. & et al. (2006), 'Inverse association between periodontitis and respiratory allergies', *Clin Exp Allergy* **36**, 495–502.

Garlet, G. (2010), 'Destructive and protective roles of cytokines in periodontitis: A re-appraisal from host defense and tissue destruction viewpoints', *Journal of Dental Research* **89**(12), 1349–1363.

Gomes-Filho, I. S., Soledade-Marques, K. R., Seixas da Cruz, S. & et al. (2013), 'Does periodontal infection have an effect on severe asthma in adults?', *Journal of Periodontology* **85**, 179–187.

Grossi, S. G., Zambon, J. J., Ho, A. W., Koch, G., Dunford, R. G., Machtei, E. E. & et al. (1994), 'Assessment of risk for periodontal disease. i. risk indicators for attachment loss.', *J Periodontol,* **65**, 260–267.

Hansel, T. T., Johnston, S. L. & Openshaw, P. J. (2013), 'Microbes and mucosal immune responses in asthma', *The Lancet* **381**, 861–873.

Hujoel, P. P., Cunha-Cruz, J., Maupome, G. & Saver, B. (2008), 'Long-term use of medications and destructive periodontal disease', *J Periodontol* **79**, 1330–1338.

| Characteristics | Asthma Prevalence | | Asthma Medication | |
|---|---|---|---|---|
| | Yes (n=191) | No (n=1,124) | Yes (n=100) | No (n=1,215) |
| Age (years) | 50.36 (6.87) | 50.5 (6.71) | 50.57 (6.90) | 50.47 (6.73) |
| Males (%) | 15 | 30 | 11 | 29 |
| Currently Smoking (%) | 23 | 19 | 23 | 19 |
| Income Level (%) | | | | |
|   < $20,000 | 63 | 55 | 65 | 56 |
|   $20,000 – $49,999 | 30 | 33 | 30 | 32 |
|   ≥ $50,000 | 7 | 12 | 5 | 12 |
| BMI (kg/m$^2$) | 34.10(6.63) | 33.19(6.17) | 34.81(6.81) | 33.20(6.18) |
| Periodontitis (%) | | | | |
|   None/Mild | 41 | 35 | 46 | 35 |
|   Moderate | 44 | 37 | 46 | 38 |
|   Severe | 15 | 28 | 8 | 27 |
| Proportion BOP* | 0.20(0.39) | 0.21(0.37) | 0.16(0.25) | 0.21(0.38) |
| Mean Plaque Index* | 0.67(0.75) | 0.67(0.83) | 0.60(0.76) | 0.67(0.83) |

Table 1: Summary of study variables conditional on asthma prevalence or on asthma medication use. Categorical variables are summarized as: % . Most quantitative variables are summarized as: mean(SD), where SD=standard deviation. Variables with * are summarized as median(IQR), where IQR= interquartile range. n is number of participants by group.

| Variables | Asthma Diagnosis | Asthma Medicine |
|---|---|---|
| CDC-APP Periodontitis | - | - |
|    None/Mild | Reference group | Reference group |
|    Moderate | 0.96 (0.68, 1.35) | 0.84 (0.54, 1.31) |
|    Severe | **0.44 (0.27, 0.70)** | **0.20 (0.09, 0.43)** |
| Proportion BOP | 0.86 (0.45, 1.62) | **0.26 (0.10, 0.67)** |
| Oral hygiene (Mean Plaque Index) | 0.84 (0.64, 1.09) | 0.76 (0.52, 1.08) |

**Table 2: OR (with 95% confidence intervals in parenthesis) for different periodontal measures adjusted for gender, smoking status (Yes, No), family asthma prevalence, income level (reference group <$20,000), and numerical variables age and BMI. Entries in bold are statistically significant.**

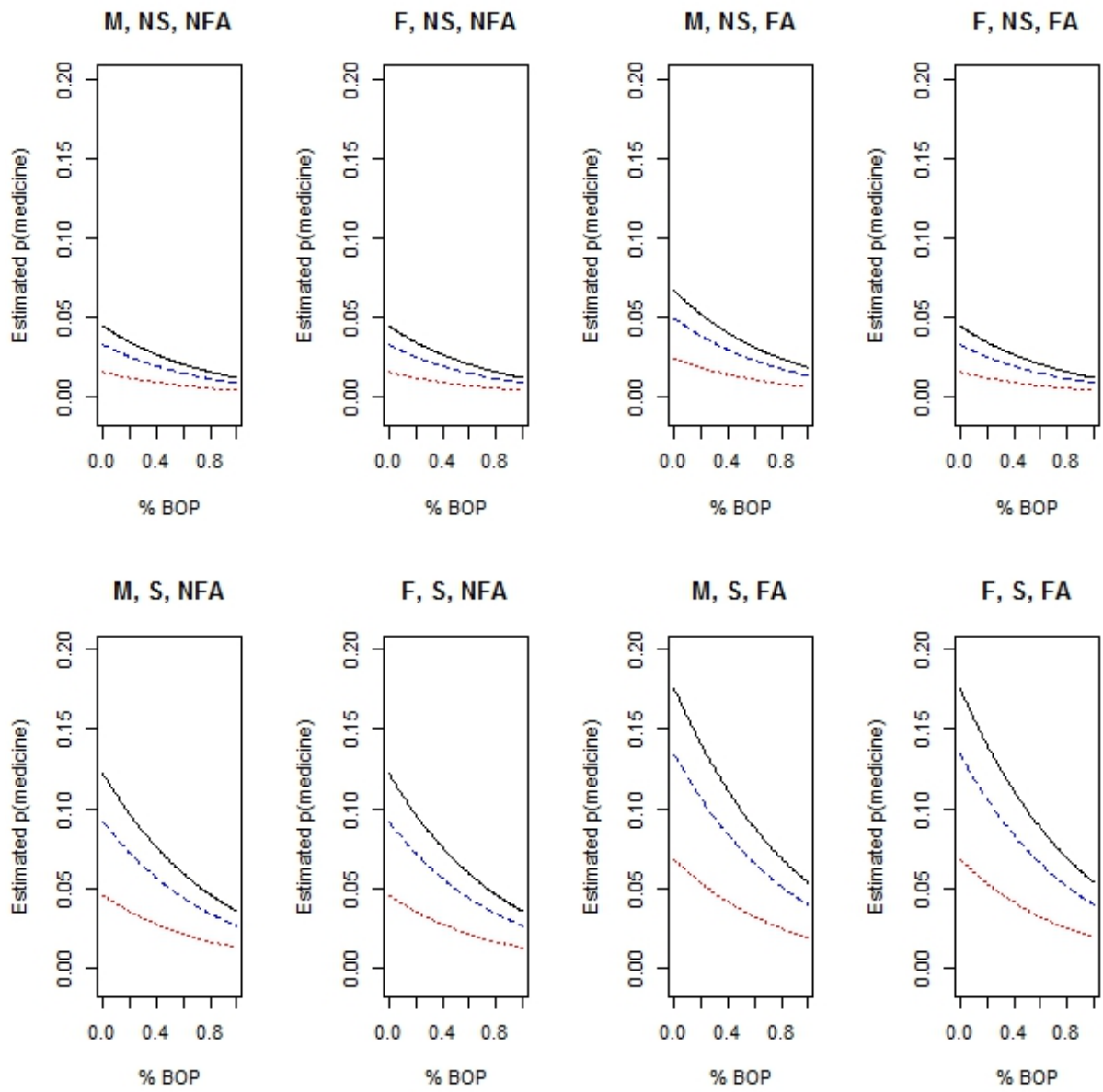

*Figure 1: Logistic regression estimating the probability of asthma medication use as a function of PBOP, according to low income (whole line), medium income (dashed line), higher income (dotted line), gender, smoking status and income level. M=male, F=Female, NS=non-smoker, S=Smoker, NFA=No family history of asthma, FA= family history of asthma. Reference age of 50 years and BMI of 33 were used*